\pdfoutput=1
\documentclass[useAMS,usenatbib,referee]{biom}
\def\bSig\mathbf{\Sigma}

\usepackage[figuresright]{rotating}
\usepackage{multicol, multirow}
\usepackage{amsmath}
\usepackage{natbib}
\usepackage{hyperref}
\usepackage{graphicx}
\usepackage{subfig}

\DeclareUnicodeCharacter{2061}{}
\DeclareUnicodeCharacter{FF0C}{}

\title[Sensitivity analysis for publication bias in meta-analysis of diagnostic studies]{A likelihood-based sensitivity analysis for addressing publication bias in meta-analysis of diagnostic studies using exact likelihood}

\author{Taojun Hu\\
	 Department of Biomedical Statistics, Graduate School of Medicine, Osaka University, Osaka, Japan\\
 Department of Biostatistics, School of Public Health, Peking University, Beijing, China\\
	 \and 
	 Yi Zhou\\
	 Beijing International Center for Mathematical Research, Peking University, Beijing, China\\
 Department of Biomedical Statistics, Graduate School of Medicine, Osaka University, Osaka, Japan
 \and
 Xiao-Hua Zhou\\
 Beijing International Center for Mathematical Research, Peking University, Beijing, China\\
 Department of Biostatistics, School of Public Health, Peking University, Beijing, China\\
 \and
 Satoshi Hattori$^{*}$\email{hattoris@biostat.med.osaka-u.ac.jp}\\
	 Department of Biomedical Statistics, Graduate School of Medicine, Osaka University, Osaka, Japan\\
 Integrated Frontier Research for Medical Science Division, \\Institute for Open and Transdisciplinary Research Initiatives (OTRI), Osaka University, Osaka, Japan\\
	 }

\begin{document}

% \date{{\it Received October} 2004. {\it Revised February} 2005.\newline 
% {\it Accepted March} 2005.}

% \pagerange{\pageref{firstpage}--\pageref{lastpage}} \pubyear{2006}

% \volume{59}
% \artmonth{December}
% \doi{10.1111/j.1541-0420.2005.00454.x}

% This label and the label ``lastpage'' are used by the \pagerange
% command above to give the page range for the article

\label{firstpage}

% pub the summary here

\begin{abstract}
Publication bias (PB) poses a significant threat to meta-analysis, as studies yielding notable results are more likely to be published in scientific journals. Sensitivity analysis provides a flexible method to address PB and to examine the impact of unpublished studies. A selection model based on t-statistics to sensitivity analysis is proposed by Copas. This t-statistics selection model is interpretable and enables the modeling of biased publication sampling across studies, as indicated by the asymmetry in the funnel-plot. In meta-analysis of diagnostic studies, the summary receiver operating characteristic curve is an essential tool for synthesizing the bivariate outcomes of sensitivity and specificity reported by individual studies. Previous studies address PB upon the bivariate normal model but these methods rely on the normal approximation for the empirical logit-transformed sensitivity and specificity, which is not suitable for sparse data scenarios. Compared to the bivariate normal model, the bivariate binomial model which replaces the normal approximation in the within-study model with the exact within-study model has better finite sample properties. In this study, we applied the Copas t-statistics selection model to the meta-analysis of diagnostic studies using the bivariate binomial model. To our knowledge, this is the first study to apply the Copas t-statistics selection model to the bivariate binomial model. We have evaluated our proposed method through several real-world meta-analyses of diagnostic studies and simulation studies.
\end{abstract}

\begin{keywords}
Diagnostic studies; Meta-analysis; Publication bias; Sensitivity analysis; Summary receiver operating characteristic
\end{keywords}

\maketitle

\section{Introduction}
\label{ss:intro}

Diagnostic studies have long been utilized to assess the effectiveness of medical tests. These tests aim to determine the presence or absence of certain diseases in patients. The Receiver Operating Characteristic (ROC) curve and the area under the ROC curve (AUC) offer thorough evaluations for diagnostic studies involving continuous biomarkers. Meta-analysis plays an important role in synthesizing the various results across multiple studies and providing a comprehensive understanding of the effectiveness of the diagnostic test. Suppose we are interested in the meta-analysis of diagnostic studies. If we can extract the AUCs from all the studies, we can follow the standard technique for univariate meta-analysis. However, some studies do not report the AUCs and instead report empirical sensitivities and specificities. Since the cut-off values to define the sensitivity and the specificity are heterogeneous over studies, the two metrics are likely to correlate with each other negatively. Thus, summarizing sensitivity and specificity independently would be less appealing.

The Summary Receiver Operating Characteristic (SROC) curve is an invaluable tool for consolidating sensitivities and specificities from various studies \citep{moses1993combining,rutter2001hierarchical,reitsma2005bivariate}. It portrays the relationship between pairs of sensitivity and 1-specificity over the individual studies. The area under the SROC curve (SAUC) serves as a useful metric to evaluate the performance of diagnostic tests. The SROC curve and SAUC are used when we do not have the AUCs for all the studies and instead, have pairs of sensitivity and specificity for all the studies. There are two primary methods to estimate the SROC curve/SAUC. The first one is the bivariate normal model \citep{reitsma2005bivariate}, which assumes that the logit-transformed empirical sensitivity and specificity follow a joint normal distribution, relying on the central limit theorem for asymptotic normality. The other is based on the bivariate binomial model. Bayesian inference \citep{rutter2001hierarchical} and the maximum likelihood method \citep{macaskill2004empirical} can be used for making inference with the bivariate binomial model. 
Since the bivariate binomial model does not rely on the asymptotic normality of the logit-transformed empirical sensitivity and specificity in constructing the likelihood function, it enhances finite-sample performance comparing to the bivariate normal model. The bivariate binomial model especially addresses challenges presented by sparse data. Sparse data situations arise in diagnostic studies with extremely low or zero occurrences of certain outcomes, especially when the true sensitivity or specificity is high in a study with a small sample size. In the presence of zero frequencies, the logit-transformed empirical sensitivity and specificity are not defined. To handle this situation with the bivariate normal model, a continuity correction is necessary and may introduce bias in inference \citep{stijnen2010random}. The bivariate binomial model avoids this issue, rendering itself more reliable in the presence of sparse data. Due to these advantages, we focus on the bivariate binomial model in this paper.

Publication bias (PB) is a critical issue that jeopardizes the reliability of meta-analyses. Studies with remarkable findings or large sample sizes are preferentially published, leading to an overrepresentation of positive findings. PB has been extensively studied in the standard meta-analysis for intervention studies. Graphical methods like the funnel-plot and trim-and-fill methods \citep{egger1997bias,duval2000trim} presented accessible ways to detect and adjust for PB. However, graphical methods could be subjective and less informative to the selection mechanism of studies. In contrast to the graphical methods, the sensitivity analysis method models the mechanism of selective publication using the selection function and would give us a more insightful interpretation of PB. \cite{copas1999works} and \cite{copas2000meta,copas2001sensitivity} first introduced a selection model based on the Heckman model \citep{heckman1976,heckman1979} and assumed whether a study would be published or not was determined by a latent Gaussian random variable. We refer to the sensitivity analysis method proposed by \cite{copas2001sensitivity} as the Copas-Heckman selection model. Modeling with a latent Gaussian random variable is convenient to handle the normal-normal (NN) random-effects model for the outcome; the likelihood conditional on published can be easily derived based on the joint normal distribution. On the other hand, since the selective publication is described through a latent variable, it cannot clarify what kind of selective publication is behind the meta-analysis. \cite{copas2013likelihood} introduced an alternative way to derive the likelihood function conditional on published by linking the distribution defined by the NN model and the observed distribution of the published studies. The method allows us to handle selection functions monotone with the t-statistic or equivalently its p-value given in each individual study.

In contrast, methods for addressing PB in meta-analysis of diagnostic studies remain underexplored. \cite{deeks2005performance} and \cite{burkner2014testing} proposed to apply the methods for the univariate meta-analysis such as Begg, Egger, and Machackill tests \citep{begg1994operating,egger1997bias,macaskill2001comparison} with some univariate diagnosis measure such as the log diagnostic odds ratio (lnDOR). However, none of them can directly address the impact of selective publication on the SROC curve. 
Consequently, developing methods to address PB on SROC curve/SAUC is more attractive. \cite{luo2022accounting} proposed a bivariate trim-and-fill method to simultaneously address PB in the potential correlated sensitivity and specificity. However, appropriateness of the funnel-plot symmetry in two dimensions is unclear. Alternatively, more formal approaches based on modelling the selective mechanism can be taken. \cite{piao2019copas} devised a likelihood-based method for parameter estimation with the Copas-Heckman selection model, applied to the bivariate normal model using an EM algorithm. \cite{li2021diagnostic} took a different approach to correct PB in the bivariate normal model using empirical likelihood, also relying on the Copas-Heckman selection model. 
\cite{zhou2023likelihood} introduced a versatile sensitivity analysis technique with the bivariate normal model utilizing the Copas t-statistics selection model. This approach has shown considerable promise in resolving PB using the Copas t-statistics selection model in the meta-analysis of diagnostic studies. Regarding the bivariate binomial model, \cite{hattori2018sensitivity} introduced a sensitivity analysis technique based on the Copas-Heckman selection model. Since the bivariate binomial model has better finite sample performance due to the use of the exact within-study model compared with the bivariate normal model, it may be beneficial to extend the Copas t-statistics selection model to the bivariate binomial model in meta-analysis of diagnostic studies.

In this paper, we propose a sensitivity analysis method to address PB in meta-analysis of diagnostic studies based on the Copas t-statistics selection model with the bivariate binomial model. We extend the likelihood-based sensitivity analysis method by \cite{copas2013likelihood} to the bivariate binomial model for meta-analysis of diagnostic studies. The argument in \cite{copas2013likelihood} was made under the NN random-effects model and we point it out that the steps can be implemented even for other models. We derive the conditional likelihood given published studies. On the other hand, in handling the exact likelihood with mixed effects, we need to make intensive calculations of integral when calculating the marginal selection probability in the iteration steps to maximize the likelihood. 
Our proposal assumes the binomial distribution of the within-study likelihood, resulting in increased computational complexity when computing the marginal probability. To address the computational issue, we propose approximation method for the marginal probability with high accuracy and small computational costs. To the best of our knowledge, this is the first study to employ the Copas t-statistics selection model in conjunction with the bivariate binomial model for the meta-analysis of diagnostic studies. The remainder of this article is structured as follows: Section 2 provides a concise overview of the bivariate binomial model without considering PB. Section 3 introduces the proposed sensitivity analysis method, which applies the Copas t-statistics selection model to the bivariate binomial model; we also elaborate on our inference approaches. In Section 4, we illustrate our proposed sensitivity analysis with real-world datasets. In Section 5, we further substantiate the performance of our proposal with simulation studies. Finally, Section 6 summarizes our methods as well as discusses our proposal with the previous related work.

\section{The bivariate binomial model}
\label{ss:glmms}

\subsection{Notations}
\label{ss:notations}

Let us consider a meta-analysis encompassing \(S\) published diagnostic studies. The number of subjects in the \(s\)-th~($s=1, \cdots, S$) study is represented as \(n^{(s)}\). Each study evaluates the diagnostic capacity of the common continuous biomarker to
determine the presence of the disease based on study-specific cut-off points. Without loss of generality, we assume that larger values of biomarker indicate that the subject is more likely to contract the disease. Consequently, subjects are classified into the tested positive group, which is denoted by $X_i^{(s)}=1$, if the biomarker is larger than the given cut-off point, or tested negative groups, which is denoted by $X_i^{(s)}=0$, if the biomarker is lower than that. Let the actual
disease status of the subject \(i\ (i=1, \ldots, n^{(s)})\) in the \(s\)-th study be denoted by \(D_i^{(s)}\).
We denote the number of subjects with test outcome \(X_i^{(s)} = x\) and
disease status \(D_i^{(s)} = d\) as \(N_{xd}^{(s)}\), and its realization as \(n_{xd}^{(s)}\), for \(x = 0, 1\) and \(d = 0, 1\). Thus,
each study is supposed to offer the information of a 2\(\times\)2 contingency table (see Web Appendix A). Within the contingency table, the true positive (TP) is $n_{11}^{(s)}$; the true negative (TN) is $n_{00}^{(s)}$; while the false negative (FN) and false positive (FP) is $n_{01}^{(s)}$ and $n_{10}^{(s)}$, respectively. The total numbers of subjects with and without disease are denoted by $n_{1}^{(s)}$ and $n_{0}^{(s)}$ respectively.

\subsection{Bivariate binomial model}\label{bivariate-binomial-model}

Let the study-specific TPR and FPR be denoted by \(\pi_1^{(s)} = P(X_i^{(s)}=1\mid D_i^{(s)}=1)\) and \(\pi_0^{(s)} = P(X_i^{(s)}=1\mid D_i^{(s)}=0)\). The bivariate binomial model \citep{macaskill2004empirical,rutter2001hierarchical} is defined by:

\begin{equation}
\begin{aligned}
\pi_1^{(s)} &= 1 - G\left\{-\frac{\theta+\theta^{(s)}+\left(\alpha+\alpha^{(s)}\right)/2}{\exp \left(\beta/2\right)}\right\}\\
\pi_0^{(s)} &= 1 - G\left\{-\frac{\theta+\theta^{(s)}-\left(\alpha+\alpha^{(s)}\right)/2}{\exp \left(-\beta/2\right)}\right\}\\,
\end{aligned}
\label{eq:p12}
\end{equation}
where \(G(\cdot)\) represents a known cumulative distribution function (c.d.f).
A commonly used choice for \(G(\cdot)\) is the standard logistic function, defined as \(G(x) = \frac{1}{1 + \exp(-x)}\). $\theta^{(s)}$ and $\alpha^{(s)}$ are random-effects, which are assumed to follow the bivariate normal distribution, given by

\begin{equation*}
\left(\begin{array}{c}
\theta^{(s)} \\
\alpha^{(s)}
\end{array}\right) \sim N\left(\left(\begin{array}{l}
0 \\
0
\end{array}\right),\left(\begin{array}{c}
\sigma_\theta^2, 0 \\
0, \sigma_\alpha^2
\end{array}\right)\right).
\end{equation*}

The within-study model is assumed to
follow a binomial distribution, defined as
\(N_{11}^{(s)} \sim \text{Binomial}\left(n_{1}^{(s)}, \pi_1^{(s)}\right)\)
and
\(N_{10}^{(s)} \sim \text{Binomial}\left(n_{0}^{(s)}, \pi_0^{(s)}\right)\) , respectively, conditional on $\theta^{(s)}$ and $\alpha^{(s)}$.
Consequently, the likelihood of the model, without accounting for PB, is given by

\[
L(\Theta) = \prod_{s=1}^S P\left(N_{11}^{(s)}=n_{11}^{(s)}, N_{10}^{(s)}=n_{10}^{(s)}, N_{01}^{(s)}=n_{01}^{(s)}, 
N_{00}^{(s)}=n_{00}^{(s)}\right)
\]
with
\begin{equation}
\begin{aligned}
& P\left(N_{11}^{(s)}=n_{11}^{(s)}, N_{10}^{(s)}=n_{10}^{(s)}, N_{01}^{(s)}=n_{01}^{(s)}, 
N_{00}^{(s)}=n_{00}^{(s)}\right)\\ = & \prod_{s=1}^S \int_{-\infty}^{\infty} \int_{-\infty}^{\infty} P\left(N_{11}^{(s)}=n_{11}^{(s)}, N_{10}^{(s)}=n_{10}^{(s)} \mid \theta^{(s)}, \alpha^{(s)}\right) f_\theta\left(\theta^{(s)}\right) f_\alpha\left(\alpha^{(s)}\right) d \theta^{(s)} d \alpha^{(s)} \\
= & \prod_{s=1}^S \int_{-\infty}^{\infty} \int_{-\infty}^{\infty} \prod_{d=0,1} \begin{pmatrix}n_{d}^{(s)}\\n_{1d}^{(s)} \end{pmatrix} \left\{\pi_d^{(s)}\right\}^{n_{1 d}^{(s)}}\left\{1-\pi_d^{(s)}\right\}^{n_{d}^{(s)}-n_{1 d}^{(s)}} f_\theta\left(\theta^{(s)}\right) f_\alpha\left(\alpha^{(s)}\right) d \theta^{(s)} d \alpha^{(s)}.
\end{aligned}
\label{eq:llk_bb}
\end{equation}
Denote $\Theta=(\theta, \alpha, \beta, \sigma_{\theta}, \sigma_{\alpha})$. Setting the random effects in~\eqref{eq:p12} as zero, the overall sensitivity and specificity are given by, 
\begin{equation}
 \begin{aligned}
 \mathrm{sensitivity} &= \pi_1 = 1 - G\left\{-\frac{\theta+\alpha/2}{\mathrm{exp}(\beta/2) }\right\}\\
 \mathrm{specificity} &= 1-\pi_0 = G\left\{-\frac{\theta-\alpha/2}{\mathrm{exp}(-\beta/2) }\right\}.
 \end{aligned}
 \label{eq:senspe}
\end{equation}
If we let \(x\) represent 1-specificity, then the sensitivity can be formulated as a
function of \(x\) with respect to the parameters \(\alpha\) and
\(\beta\), after eliminating \(\theta\).
Consequently, the SROC curve, which illustrates the relationship between sensitivity and 1-specificity, is given by \citep{hattori2018sensitivity}:

\begin{equation}
\operatorname{SROC}(x ; \alpha, \beta)=1-G\left\{-\alpha \exp({-{\beta}/{2}})+\mathrm{exp}(-\beta) G^{-1}(1-x)\right\}, 
\label{eq:hsroc}
\end{equation}
and the SAUC is given by:

\begin{equation}
\operatorname{SAUC}(\alpha, \beta)=\int_0^1 \operatorname{SROC}(x ; \alpha, \beta) d x.
\label{eq:SAUC}
\end{equation}

\section{Sensitivity analysis to adjust for publication bias in meta-analysis of diagnostic studies}
\label{ss:pb}

Following \cite{zhou2023likelihood}, we consider the linear combination of logit-transformed sensitivity (denoted by logit(sen)) and specificity (denoted by logit(spe)) as the key statistic in the selection model, that is 
\begin{equation}
c_0\mathrm{logit(spe)}+c_1\mathrm{logit(sen)},
 \label{eq:keystat}
\end{equation}
where $\mathrm{logit}(x)$ is defined as $\log \frac{x}{1-x}$. The lnDOR is a special case when $c_0=c_1=1$. For the $s$-th study, the empirical key statistic is given by $c_0\mathrm{log}\frac{n_{00}^{(s)}}{n_{01}^{(s)}}+c_1\mathrm{log} \frac{n_{11}^{(s)}}{n_{10}^{(s)}}$. The empirical variance for~\eqref{eq:keystat} is given by
\begin{equation*}
 c_0^2\left(\frac{1}{n_{01}^{(s)}} + \frac{1}{n_{00}^{(s)}}\right) + c_1^2\left(\frac{1}{n_{11}^{(s)}} + \frac{1}{n_{10}^{(s)}}\right)
\end{equation*}
for the $s$-th study. Thus, the t-statistic is given by 
\begin{equation}
 t^{(s)}=\frac{c_0\mathrm{log}\frac{n_{00}^{(s)}}{n_{01}^{(s)}}+c_1\mathrm{log} \frac{n_{11}^{(s)}}{n_{10}^{(s)}}}{\sqrt{c_0^2\left(\frac{1}{n_{01}^{(s)}} + \frac{1}{n_{00}^{(s)}}\right)+c_1^2\left(\frac{1}{n_{11}^{(s)}} + \frac{1}{n_{10}^{(s)}}\right)}}.
 \label{eq:tstatistics}
\end{equation}
As the t-statistics are scale-invariant, we constrain $c_1^2+c_0^2=1$ without loss of generality. The t-statistics for the lnDOR corresponds to $c_0=c_1=1/\sqrt{2}$. The proposed selection function can be expressed as follows:

\begin{equation}
 P\left(\mathrm{select}\mid \gamma_0, \gamma_1, t^{(s)}\right) = a\left(t^{(s)}\right) = H\left(\gamma_0 + \gamma_1 t^{(s)}\right),
 \label{eq:selection}
\end{equation}
where $H(\cdot)$ is an arbitrary non-decreasing function valued from 0 to 1. The probit function $\Phi(\cdot)$ is a popular choice for $H(\cdot)$ \citep{copas2013likelihood}. If not specified, the probit function $\Phi(\cdot)$ is used in this paper. When $(c_0,c_1)=(1/\sqrt{2},1/\sqrt{2})$, the selective publication is determined by the significance of the lnDOR; when $(c_0,c_1)=(0,1)$ or $(1,0)$, the selective publication is determined by the significance of sensitivity or specificity, respectively. For studies with zero cells in the contingency table, $t^{(s)}$ is undefined. We apply the continuity correction, which is adding 0.5 to all the cells, and then calculate $t^{(s)}$. We note that the continuity correction is used for study-specific t-statistic in the selection function~\eqref{eq:selection} to model the selective publication and is \textit{not} used in constructing the exact likelihood of meta-analysis. 

The likelihood conditional on published studies is given by

\begin{equation}
\begin{aligned}
\mathcal{L}_O(\Theta, \gamma_0, \gamma_1) &= \prod_{s=1}^S P(N_{11}^{(s)} = n_{11}, N_{01}^{(s)} = n_{01}, N_{10}^{(s)} = n_{10}, N_{00}^{(s)} = n_{00}\mid \mathrm{select} )\\
&= \prod_{s=1}^S \frac{P(\mathrm{select}\mid n_{11}^{(s)}, n_{10}^{(s)}, n_{01}^{(s)}, n_{00}^{(s)})f_P(n_{11}^{(s)}, n_{10}^{(s)}, n_{01}^{(s)}, n_{00}^{(s)})f_P(n_1^{(s)}, n_0^{(s)})}{P(\mathrm{select})}\\
&= \prod_{s=1}^S \frac{P(\mathrm{select}\mid \gamma_0, \gamma_1, t^{(s)})f_P(n_{11}^{(s)}, n_{10}^{(s)}, n_{01}^{(s)}, n_{00}^{(s)})f_P(n_1^{(s)}, n_0^{(s)})}{P(\mathrm{select})} \\
&= \prod_{s=1}^S \frac{a(t^{(s)})f_P(n_{11}^{(s)}, n_{10}^{(s)}, n_{01}^{(s)}, n_{00}^{(s)})f_P(n_1^{(s)}, n_0^{(s)})}{P(\mathrm{select})},
\end{aligned}
\label{eq:llk}
\end{equation}
where the $f_P(n_{11}^{(s)}, n_{10}^{(s)}, n_{01}^{(s)}, n_{00}^{(s)})$ refers to $P\left(N_{11}^{(s)}=n_{11}^{(s)}, N_{10}^{(s)}=n_{10}^{(s)}, N_{01}^{(s)}=n_{01}^{(s)}, 
N_{00}^{(s)}=n_{00}^{(s)}\right)$ in~\eqref{eq:llk_bb}, while the $f_P\left(n_1^{(s)}, n_0^{(s)}\right)$ denotes the marginal probability mass function of $\left(n_1^{(s)},n_0^{(s)}\right)$, and the denominator $P(\mathrm{select})$ denotes the marginal selection probability. In the following part, we denote the population probability or density with the suffix $P$ whereas those conditional on published studies with the suffix $O$.

Following \cite{copas2013likelihood}, we assume a fixed marginal selection probability \(P(\mathrm{select}) = p\). Given \(p\), the probability taking the value \((n_1^{(s)}, n_0^{(s)})\) conditional on published is given by

\begin{equation}
f_O(n_1^{(s)}, n_0^{(s)}) = P(n_1^{(s)}, n_0^{(s)} \mid \mathrm{select}) = \frac{P\left(\text { select } \mid n_1^{(s)}, n_0^{(s)}\right) f_P\left(n_1^{(s)}, n_0^{(s)}\right)}{p}.
 \label{eq:org}
\end{equation}
An equivalent form is
\begin{equation*}
f_P\left(n_1^{(s)}, n_0^{(s)}\right)=p \frac{1}{P\left(\text { select } \mid n_1^{(s)}, n_0^{(s)}\right)} f_O\left(n_1^{(s)}, n_0^{(s)}\right).
\end{equation*}
Integrating it over \(\left(n_1^{(s)}, n_0^{(s)}\right)\) in both sides, we can derive
\begin{equation}
\frac{1}{p}=E_O\left(\frac{1}{P\left(\text { select } \mid n_1^{(s)}, n_0^{(s)}\right)}\right),
\label{eq:restriction}
\end{equation}
where $E_O$ means the expectation conditional on published studies. We consider the empirical version of~\eqref{eq:restriction} by replacing the expectation conditional on published with the average over observed studies, that is
\begin{equation}
 \dfrac{1}{p}= \frac{1}{S}\sum_{s=1}^{S}P\left(\text { select } \mid n_1^{(s)}, n_0^{(s)}\right).
 \label{eq:constraint}
\end{equation}

Combining with~\eqref{eq:org}, the likelihood conditional on published~\eqref{eq:llk} can be rewritten as

\begin{equation*}
\begin{aligned}
\mathcal{L}_O(\Theta, \gamma_0, \gamma_1) &= \prod_{s=1}^S \frac{ a(t^{(s)}) f_P(n_{11}^{(s)}, n_{10}^{(s)}, n_{01}^{(s)}, n_{00}^{(s)})f_P(n_1^{(s)}, n_0^{(s)}) f_O(n_1^{(s)}, n_0^{(s)})}{P(\mathrm{select}\mid n_1^{(s)}, n_0^{(s)}) f_P(n_1^{(s)}, n_0^{(s)}) }\\
&= \prod_{s=1}^S \frac{a(t^{(s)}) f_P(n_{11}^{(s)}, n_{10}^{(s)}, n_{01}^{(s)}, n_{00}^{(s)}) f_O(n_1^{(s)}, n_0^{(s)})}{P(\mathrm{select}\mid n_1^{(s)}, n_0^{(s)}) }.
\end{aligned}
\end{equation*}

Thus we can derive the log-likelihood conditional on published, that is,

\begin{equation}
\begin{aligned}
\ell_O (\Theta, \gamma_0, \gamma_1)&=\sum_{s=1}^S \log f_P\left(n_{11}^{(s)}, n_{10}^{(s)}, n_{01}^{(s)}, n_{00}^{(s)}\right)+\sum_{s=1}^S \log a\left(t^{(s)} \right)-\\
&\sum_{s=1}^S \log P\left(\text { select } \mid n_1^{(s)}, n_0^{(s)}\right)+\sum_{s=1}^S \log f_O\left(n_1^{(s)},n_0^{(s)} \right). 
\end{aligned}
\label{eq:log-likelihood}
\end{equation}

Equation~\eqref{eq:constraint} poses a constraint to the likelihood. We estimate the parameters $(\Theta, \gamma_0, \gamma_1)$ and make inferences on them by maximizing the log-likelihood~\eqref{eq:log-likelihood} with the constraint function~\eqref{eq:constraint}. On the right side of~\eqref{eq:log-likelihood}, the last term is a constant free from the parameters. The first and third terms are given by~\eqref{eq:llk_bb} and~\eqref{eq:selection}, respectively. The major problem lies in
the calculation for the \(P\left(\text{select} \mid n_1^{(s)}, n_0^{(s)}\right)\) in the third term of~\eqref{eq:log-likelihood} and~\eqref{eq:constraint}. It is represented by
\begin{equation}
\begin{aligned}
&P\left(\text { select } \mid n_1^{(s)}, n_0^{(s)}\right) \\=& \sum_{m_{11}=0}^{n_1^{(s)}}\sum_{m_{00}=0}^{n_0^{(s)}}P\left(\text { select } \mid m_{11}, n_{1}^{(s)}, m_{00}, n_{0}^{(s)}\right) f_P(m_{11}, n_1^{(s)}- m_{11}, n_0^{(s)}-n_{01}^{(s)}, m_{00})\\
=& \sum_{m_{11}=0}^{n_1^{(s)}}\sum_{m_{00}=0}^{n_0^{(s)}}a\left(t(m_{11}, n_{1}^{(s)}, m_{00}, n_{0}^{(s)})\right)f_P\left(m_{11}, n_1^{(s)}- m_{11}, n_0^{(s)}-n_{01}^{(s)}, m_{00}\right),
\end{aligned}
\label{eq.summing}
\end{equation}
where
\begin{equation*}
t(m_{11}, n_{1}^{(s)}, m_{00}, n_{0}^{(s)}) = \frac{c_1\mathrm{log} \frac{m_{11}}{n_1^{(s)}- m_{11}}+c_0\mathrm{log}\frac{m_{00}}{n_0^{(s)}-m_{00}}}{\sqrt{c_1^2\left(\frac{1}{m_{11}} + \frac{1}{n_1^{(s)} -m_{11}}\right) + c_0^2\left(\frac{1}{n_0^{(s)} -m_{00}} + \frac{1}{m_{00}}\right)}}.
 % \label{eq.add1}
\end{equation*}
For simplification, we denote $t(m_{11}, n_{1}^{(s)}, m_{00}, n_{0}^{(s)})$ as $\Tilde{t}^{(s)}(m_{11}, m_{00})$. As given in~\eqref{eq:llk_bb}, \(f_P(\cdot)\) involves bivariate integration.
We need to calculate many integrals in~\eqref{eq.summing} and sum them up. It is computationally demanding. In addition to the computational complexity, we may suffer from the issue of error accumulation. 
We address the problem through an approximation approach. We see $m_{11}, m_{00}$ as random variables following binomial distributions, that is, $m_{11} \sim \mathrm{Binomial}\left(n_1^{(s)}, p_1^{(s)}\right)$ and $m_{00} \sim \mathrm{Binomial}\left(n_0^{(s)}, p_0^{(s)}\right)$. Then 
$P(\mathrm{select}\mid n_1^{(s)},n_0^{(s)}) = E_{\Tilde{t}^{(s)}(m_{11}, m_{00})}H\left(\gamma_0+\gamma_1 \Tilde{t}^{(s)}(m_{11}, m_{00}))\right)$, where $E_{\Tilde{t}^{(s)}(m_{11}, m_{00}))}$ implies the expectation with respect to the distribution of $\Tilde{t}^{(s)}(m_{11}, m_{00}))$.
We use the asymptotic normality properties of the statistics $\Tilde{t}^{(s)}(m_{11}, m_{00}))$ to approximate its true distribution and then derive an approximation for $P(\mathrm{select}\mid n_1^{(s)},n_0^{(s)})$. While the direct calculation for~\eqref{eq.summing} needs continuity correction to compute $\Tilde{t}^{(s)}(m_{11}, m_{00}))$ if any of $m_{11}, n_1^{(s)}-m_{11}, m_{00}$ or $ n_{0}^{(s)}- m_{00}$ is zero, our approximation method avoids using the continuity correction. The proof of the asymptomatic normality for $\Tilde{t}^{(s)}(m_{11}, m_{00}))$ and the derivation of its asymptomatic expectation and variance are placed in the Web Appendix B.

Following \cite{copas2013likelihood}, we can derive the
parameter \(\gamma_0\) as a function of the remaining parameters based on the constraint function~\eqref{eq:constraint}, denoted by
\( \widehat{\gamma}_0 = \widehat{\gamma}_0(\Theta, \gamma_1)\). Thus, the log-likelihood conditional on published can be determined by \((\Theta, \gamma_1)\). The
maximum likelihood estimators (MLE) of the parameters are denoted by \( \widehat{\Theta}, \widehat{\gamma}_1\). We
estimate the asymptomatic variance matrix for the parameters \(\Theta\)
through the inverse of observed Fisher information, which is denoted by \( \widehat{\Sigma}\).
The SAUC can be estimated through the plug-in of
\( \widehat{\Theta}\) to the expressions of SAUC~\eqref{eq:SAUC}, and then the resulting estimator is denoted by
\(\widehat{SAUC} = SAUC( \widehat{\alpha}, \widehat{\beta})\). The
variance of the SAUC can be estimated through the Delta method, that is

\begin{equation*}
\begin{aligned}
\mathrm{Var}(\widehat{SAUC}) &= \frac{\partial SAUC}{\partial (\alpha, \beta)}\mid _{ \widehat{\alpha}, \widehat{\beta}} \widehat{\Sigma}_{{\alpha}, {\beta}} \frac{\partial SAUC}{\partial (\alpha, \beta)}\mid _{ \widehat{\alpha}, \widehat{\beta}}^{T},
\end{aligned}
\end{equation*}
where the superscript \(T\) means the transpose of the vector and the
\( \widehat{\Sigma}_{{\alpha}, {\beta}}\) is the submatrix of $ \widehat{\Sigma}$ corresponding to the components for $(\alpha,\beta)$.

We construct a 95\% confidence interval (CI), which is always within [0,1] by applying the Delta method with some transformation $g$ from (0,1) to $(-\infty, \infty)$. The asymptomatic expectation and variance of $g$-transformed SAUC are given by $g(\widehat{SAUC})$ and $\{g^{\prime}(\widehat{SAUC})\}^2\mathrm{Var}(\widehat{SAUC})$, where $g^{\prime}=\mathrm{d} g(x)/\mathrm{d}x$. The reconstructed 95\% CI for SAUC can be expressed as

\begin{equation*}
\begin{aligned}
g^{-1}\left\{g(\widehat{SAUC})\right. &\left.\pm 1.96 g^{\prime}(\widehat{SAUC})\sqrt{\mathrm{Var}(\widehat{S A U C})}\right\}.\\
\end{aligned}
\end{equation*}
A common choice for $g$ is the logit function, $g(x) = \mathrm{log} \frac{x}{1-x}$. We then vary the marginal selection probabilities $p$ to view how PB would affect the estimate for the parameters in the model and the SAUC.

\section{Application}
\label{ss:realanalysis}

We illustrated our sensitivity analysis method with the meta-analysis conducted by \cite{li2013neutrophil}. \cite{zhou2023likelihood} used this dataset for their illustration, and the issue of sparsity was overlooked in their analysis. This meta-analysis evaluated the effectiveness of the neutrophil CD64 expression as an biomarker to differentiate infected patients from non-infected ones with bacterial infection with 27 individual studies. The data is presented in Table~\ref{tab:cd64}. Of the 27 independent studies, two studies (No.7 and 20) had zero frequencies of FP, and most of the other studies had very low frequencies of FP or FN. In the original paper, \cite{li2013neutrophil} estimated the SAUC as 0.925, highlighting the potential of neutrophil CD64 expression as a promising biomarker for diagnosing bacterial infection. It pointed out that the meta-analysis suffered from great PB since Egger's test showed a remarkable trend of PB ($p<0.001$). We re-analyzed this meta-analysis data, whose original paper did not address how PB could affect the estimate of the SAUC. We conducted sensitivity analysis setting $p = 0.2, 0.4, 0.6, 0.8, 1$ with three specified selective mechanisms: $(c_0, c_1)=(1/\sqrt{2}, 1/\sqrt{2})$, $(c_1, c_0)=(1, 0)$, and $(c_1, c_0)=(0, 1)$. Note that $p=1$ implies the original analysis using the bivariate binomial model without accounting for selective publication. Within the sensitivity analysis, the link function $G(\cdot)$ in~\eqref{eq:p12} is chosen as the standard logistic function. 

\begin{table}
 \centering
 \caption{Data of meta-analysis of CD64 study}
 \label{tab:cd64}
 \begin{tabular}{ccccccc}
 \hline Study & Author & TP & FP & FN & TN & Cut-off points \\
 \hline 1 & Icardi & 53 & 6 & 3 & 47 & 1.19 \\
 2 & Cid & 100 & 10 & 15 & 7 & 1.5 \\
 3 & Gamez-Diaz & 266 & 73 & 138 & 133 & 1.7 \\
 4 & Groselj-Gren & 13 & 6 & 4 & 23 & 1.86 \\
 5 & Gros & 91 & 16 & 54 & 132 & 2.2 \\
 6 & Bhandari & 89 & 63 & 39 & 102 & 2.3 \\
 7 & Groselj-Gren & 17 & 0 & 7 & 12 & 2.38 \\
 8 & Groselj-Gren & 19 & 3 & 10 & 24 & 2.45 \\
 9 & Dilli & 31 & 6 & 4 & 36 & 4.39 \\
 10 & Genel & 40 & 8 & 9 & 27 & 3.05MFI \\
 11 & Tang & 50 & 10 & 14 & 32 & 8.5MFI \\
 12 & Hussein & 17 & 2 & 1 & 16 & 43.5MFI \\
 13 & Mokuda & 14 & 2 & 1 & 23 & 1800mol \\
 14 & Nishino & 19 & 2 & 6 & 34 & 2000mol \\
 15 & Doi & 19 & 1 & 12 & 67 & 2000mol \\
 16 & Tanaka & 28 & 2 & 18 & 93 & 2000mol \\
 17 & Matsui & 51 & 7 & 4 & 195 & 2000mol \\
 18 & Allen & 23 & 4 & 4 & 40 & 2000mol \\
 19 & Cardelli & 50 & 3 & 2 & 57 & 2398mol \\
 20 & Livaditi & 35 & 0 & 2 & 10 & 2566mol \\
 21 & Ng & 30 & 7 & 2 & 51 & 4000mol \\
 22 & Hsu & 49 & 1 & 6 & 10 & 4300mol \\
 23 & Lam & 107 & 37 & 29 & 137 & 6010mol \\
 24 & Ng & 72 & 20 & 21 & 175 & 6136mol \\
 25 & Ng & 91 & 25 & 24 & 198 & 6136mol \\
 26 & Tillinger & 21 & 2 & 1 & 74 & 10000mol \\
 27 & Layseca-Esp & 8 & 1 & 23 & 16 & \\
 \hline
 \end{tabular}
\end{table}

\begin{figure}
 \centering
 \includegraphics[width=0.95\textwidth]{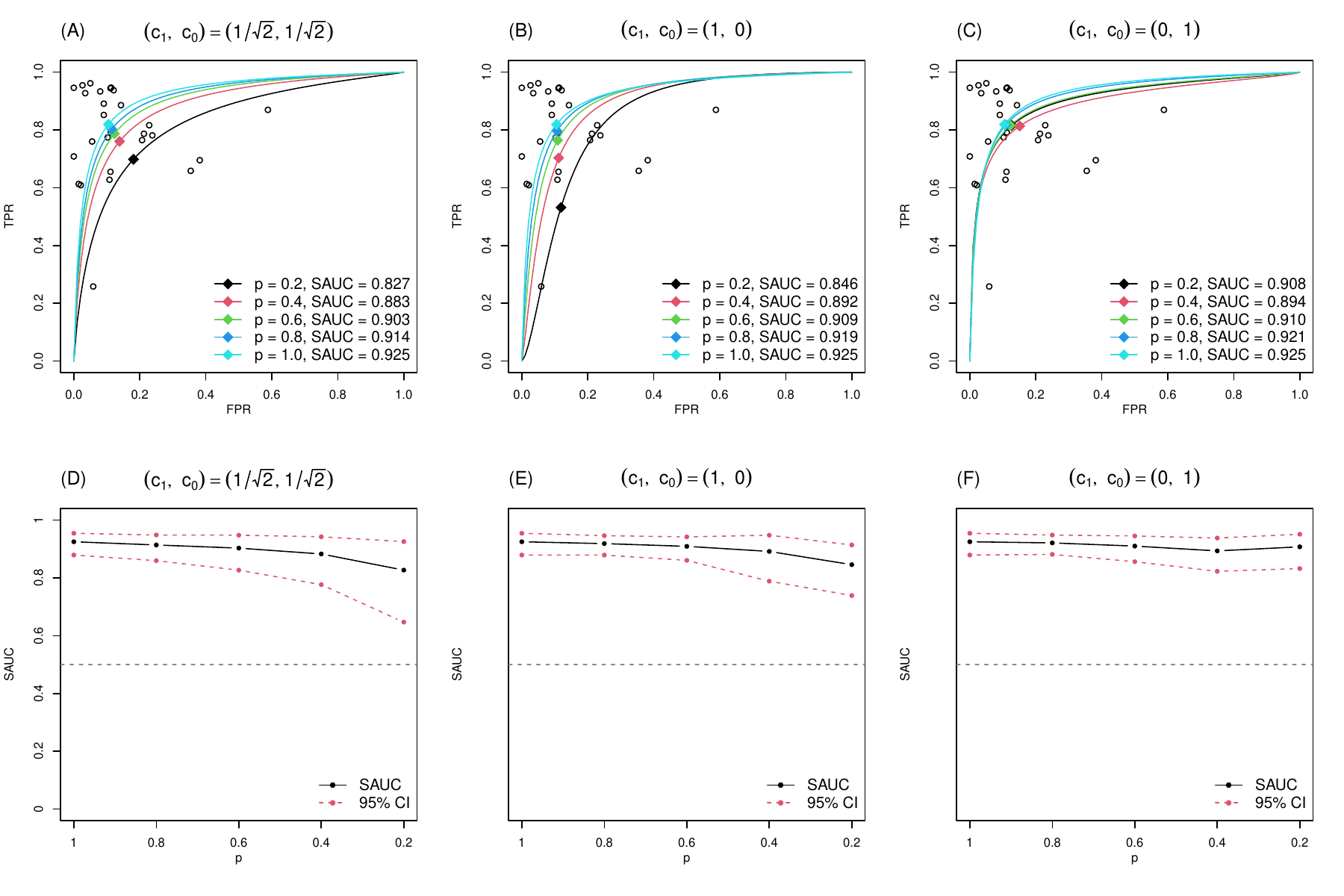}
 \caption{The estimated SROC curves and SAUC under three selective publication mechanisms in CD64 example with the bivariate binomial model (our proposal).}
 \label{fig:cd64}
\end{figure}

In the upper panels of Figure~\ref{fig:cd64}, we depicted the estimated SROC curves under various $p$ with the three settings of $(c_1, c_0)$.
The summary operating point (SOP) is the pair of the overall sensitivity and 1-specificity in~\eqref{eq:senspe} with bivariate model. The SOP is obtained by plugging the estimated parameters of the bivariate binomial model into~\eqref{eq:senspe}. It may be used as a summary measure of diagnostic capacity. However, it depends on the cut-off values of the studies in the meta-analysis. Due to this dependence of the cut-off values, the SOP is less appealing as a measure of diagnostic capacity than the SROC curve/SAUC. On the other hand, \cite{zhou2023likelihood} presented that tracing the SOP over $p$ was useful to characterize what kind of selective process was considered with the supposed selection function. Thus, we also present the SOPs for this purpose. As shown in the panel (A) of Figure~\ref{fig:cd64}, the plots of the SOPs suggested 
% that under the selection mechanism as $c_0=c_1$, 
a selective publication process under which studies around the lower right part of the SROC curve were less likely published. The change of SROC curve and SAUC suggested that the test accuracy was was robust against the selective publication mechanism determined by the significance of the lnDOR. Tracing the SOPs in the panels (B) to (C) of Figure~\ref{fig:cd64}, one could understand that the selection function with $(c_1,c_0)=(1, 0)$ and $=(0, 1$) modeled different publication mechanisms and the figures indicated that impacts by the selective publication mechanisms determined by sensitivity and specificity would be minor. 
We showed the estimated SAUC with varying $p$ in panels (D) to (F) of Figure~\ref{fig:cd64}. The SAUC was 0.925 (95\% CI: [0.880, 0.954]) without accounting for selective publication ($p=1$). With all specified values of $p$, the lower bound for SAUCs were larger than 0.5 under all three selective mechanisms, suggesting the effectiveness of neutrophil CD64 expression in diagnosing bacterial infection since SAUC=0.5 indicated that the diagnostic test result was a random guess. Of the three selective mechanisms, the estimated SAUC under $c_0=c_1$ showed a larger difference between $p=1$ and $p=0.2$ compared with the rest two selective mechanisms, suggesting that considerable PB would exist if both the sensitivity and specificity were affecting the selection. When assuming $c_0=c_1$, the SAUC under $p=0.2$ would be 0.1 lower than that under $p=1$, showing PB largely affected the estimate of SAUC. 

\begin{figure}
 \centering
 \includegraphics[width=0.95\textwidth]{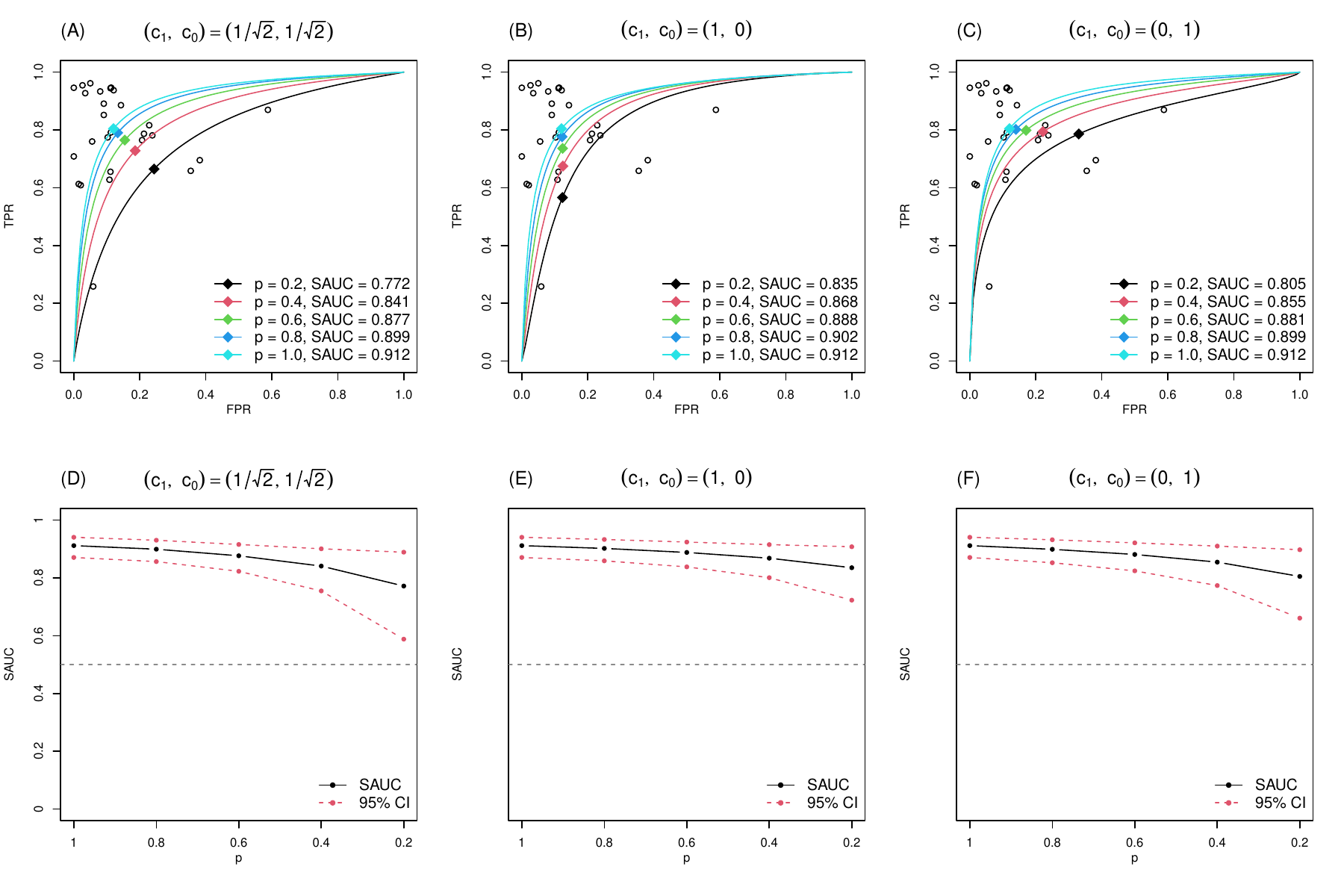}
% \caption{The estimated SROC curves and SAUC under three selective publication mechanisms in CD64 example with the bivariate normal model by \cite{zhou2023likelihood}.}
 \caption{The estimated SROC curves and SAUC under three selective publication mechanisms in CD64 example with the bivariate normal model.}
 \label{fig:cd64_nn}
\end{figure}

We also conducted the sensitivity analysis based on the bivariate normal model by \cite{zhou2023likelihood} for comparisons. 
Although the parameterizations between the bivariate binomial and bivariate normal models are different, \cite{harbord2007unification} showed there was a correspondence between the parameters of the bivariate normal model and those of the bivariate binomial model. Each model has its definition of the SROC curve and SAUC; the bivariate binomial model naturally leads to the SROC curve~\eqref{eq:hsroc} and SAUC~\eqref{eq:SAUC}, whereas \cite{reitsma2005bivariate} introduced an alternative definition of the SROC with the bivariate normal model. With the correspondence between the parameters of the two models, one may derive the SROC curve/SAUC with~\eqref{eq:hsroc} and~\eqref{eq:SAUC} even if the bivariate normal model is used for parameter estimation. The formula to this end is given in the equation (A1) of \cite{zhou2023likelihood}. The estimated SROC curves with the bivariate normal model were given in panels (A) to (C) of Figure~\ref{fig:cd64_nn} assuming different selective mechanisms under $p = 0.2, 0.4, 0.6, 0.8, 1$; the variation of SAUC was shown in panels (D) to (F) of Figure~\ref{fig:cd64_nn}. We could observe that the bivariate normal model obtained lower estimates for SAUC compared with the bivariate binomial model.

We also analyzed two additional meta-analysis datasets of diagnostic studies to show the wide applicability of our proposal. The results were placed in the Web Appendix C and D. 

\section{Simulation studies}
\label{ss:simu}

We evaluated the performance of the proposed methods with simulation studies. We considered three sets of overall sensitivity and specificity in~\eqref{eq:senspe}: (0.9, 0.5), (0.5, 0.9) and (0.8, 0.8) given the link function $G(\cdot)$ as the standard logistic function. With the scale parameter $\beta=0.15$ set, we determined three settings of $\theta$ and $\alpha$ by solving~\eqref{eq:senspe}. We set the standard deviation of cut-off parameters and accuracy parameters $(\sigma_{\theta}, \sigma_{\alpha})=(0.6, 1.2), (1.2, 0.6)$ to see if the performance of our proposal would be affected by varying the variance parameters. Thus, we totally considered 6 scenarios for data-generation and the parameters are summarized in the Web Appendix E. We considered simulating small-scale, medium-scale, and relatively large-scale meta-analyses with $\Tilde{S}\ (\Tilde{S}=15, 25, 50)$ published and unpublished studies.

Under each scenario, we generated 100 independent meta-analyses with $\Tilde{S}$ studies. For each dataset, the number of subjects with the disease $n_{1}^{(s)}$ was drawn from a discrete uniform distribution $U(10, 30)$, and the number of subjects without the disease $n_{0}^{(s)}$ drawn from the discrete uniform distribution $U(200, 300)$, mirroring real-world disparities in subject numbers. The variations in cut-off and accuracy parameters across studies, $\theta^{(s)}$ and $\alpha^{(s)}$, were sampled from normal distributions $N(0, \sigma_{\theta}^2)$ and $N(0, \sigma_{\alpha}^2)$, respectively. Then we set $\pi_1^{(s)}$ and $\pi_0^{(s)}$ with~\eqref{eq:p12}. $n_{11}^{(s)}$ and $n_{10}^{(s)}$ were generated from binomial distributions $\mathrm{Binomial}(n_1^{(s)}, \pi_1^{(s)} )$ and $\mathrm{Binomial}(n_0^{(s)}, \pi_0^{(s)} )$, respectively.

We considered the selection function based on the t-statistics for the lnDOR, which corresponded to $(c_1, c_0)=(1/\sqrt{2},1/\sqrt{2})$ in~\eqref{eq:tstatistics}. We show the results with simulation datasets under the true selective mechanisms as $(c_0, c_1)=(1, 0)$ and $(0, 1)$ in Web Appendix F. With the selection parameter $\gamma_1$ set at 1.5, we sought for $\gamma_0$ satisfying the $1/\Tilde{S}\sum_{s=1}^{\Tilde{S}}P(\mathrm{select}\mid n_{i1}^{(s)}, n_{i0}^{(s)})$ given the fixed marginal publication probability $p$ as 0.7, indicating that approximately 70\% of studies would be published. We further decided whether an individual study would be published or not by a Bernoulli distribution, $\mathrm{Bern}(\Phi(\gamma_0 + \gamma_1t^{(s)}))$.

To confirm whether the generated dataset had sparsity, we evaluated the proportion of the studies with zero entries in the $2 \times 2$ contingency table among all the studies (published and unpublished), as well as among the published studies. In addition, we evaluated the proportion of the studies with cell frequencies of no more than 3 and no more than 5. Their averages over the 100 simulated datasets are summarized in Table~\ref{tab:sparsity}. It indicates that we successfully generated meta-analyses of sparsity.

\begin{table}[]
 \centering
 \caption{Summary of the sparsity of the simulated datasets. Full indicates published and unpublished studies; Published indicates published studies.}
 \label{tab:sparsity}
 \resizebox{\linewidth}{!}{
  \begin{tabular}{cccccccc}
 \hline \multirow{2}{*}{ Experiment } & \multirow{2}{*}{ Rate (\%) } & \multicolumn{2}{c}{$S=15$} & \multicolumn{2}{c}{$S=25$} & \multicolumn{2}{c}{$S=50$} \\
 \cline{3-8} & & Full & Published & Full & Published & Full & Published \\
 \hline \multirow{3}{*}{1} & Zero entries & 19.6 & 19.6 & 20.6 & 22.1 & 19.2 & 20.4 \\
 & No-more-than-3-entries & 76 & 82.4 & 76 & 80.8 & 74.7 & 80.1 \\
 & No-more-than-5-entries & 89.8 & 93.8 & 90.8 & 93.5 & 90.1 & 94 \\
 \hline \multirow{3}{*}{2} & Zero entries & 1 & 0.8 & 0.8 & 0.7 & 1 & 0.6 \\
 & No-more-than-3-entries & 17.7 & 13.4 & 18.2 & 14 & 18.1 & 13.8 \\
 & No-more-than-5-entries & 42.8 & 36.8 & 42.8 & 37 & 41.9 & 36.1 \\
 \hline \multirow{3}{*}{3} & Zero entries & 6.7 & 4.7 & 7.2 & 5.6 & 6.8 & 6 \\
 & No-more-than-3-entries & 47.8 & 49.3 & 47.7 & 50.8 & 46.3 & 49.4 \\
 & No-more-than-5-entries & 70.8 & 73.9 & 69.8 & 73.5 & 69.3 & 72.5 \\
 \hline \multirow{3}{*}{4} & Zero entries & 23.8 & 15.7 & 23.8 & 16.3 & 23.4 & 15 \\
 & No-more-than-3-entries & 75.2 & 68.8 & 71 & 63.3 & 70 & 62.7 \\
 & No-more-than-5-entries & 87.3 & 83.3 & 86 & 81.8 & 84.9 & 80.5 \\
 \hline \multirow{3}{*}{5} & Zero entries & 3.3 & 1.4 & 3 & 1 & 3.7 & 1.1 \\
 & No-more-than-3-entries & 30.7 & 21.4 & 29.5 & 20.3 & 29.9 & 21.4 \\
 & No-more-than-5-entries & 53.9 & 45.9 & 53.2 & 44.2 & 53 & 44.7 \\
 \hline \multirow{3}{*}{6} & Zero entries & 9.8 & 4 & 10.5 & 3.9 & 11.1 & 3.5 \\
 & No-more-than-3-entries & 50.4 & 40.2 & 51 & 40.2 & 51.9 & 40.6 \\
 & No-more-than-5-entries & 70.7 & 63 & 70.1 & 62.4 & 71.6 & 63.8 \\
 \hline
 \end{tabular}
 
 }
\end{table} 

We applied our proposed method to only published studies and compared our proposal with the sensitivity analysis proposed by \cite{zhou2023likelihood} based on the bivariate normal model. Specifying $p=0.7$ and the t-statistics with $(c_1, c_0)=(1/\sqrt{2},1/\sqrt{2})$ in~\eqref{eq:tstatistics}, we summarized the estimates for parameters in the model and the SAUC in Table~\ref{tab:ressimulate}. 
The average for the SAUC by maximizing the likelihood with the bivariate binomial model with only published study (denoted as the MLE with published studies) had a non-ignorable discrepancy with the true value, indicating that PB was a considerable issue for this dataset. Both the proposed method and the method by \cite{zhou2023likelihood} successfully reduced biases. Our proposed method had smaller biases than the method by \cite{zhou2023likelihood}. We also showed the estimated SAUC with two misspecified t-statistics in the selection function~\eqref{eq:tstatistics} corresponding to $(c_1, c_0)=(1, 0)$ and $(0,1)$ in Table~\ref{tab:ressimulate}. 
% The results with $(c_1, c_0)=(1, 0)$ corresponded to estimates which adjusted for PB with only sensitivity and the results with $(c_1, c_0)=(1, 0)$ corresponded to that with only specificity. 
We observed that the misspecification of the t-statistics in the selection function could lead to slightly larger bias compared with that under the correct specified t-statistics. Correct specification of the t-statistics in the selection function could remove the bias. Except for the estimates of SAUC, we showed the estimates of $(\theta, \alpha)$ in the bivariate binomial model and the pairs of sensitivity and specificity in Web Table 5. It substantiated that our proposed method could also obtain the least bias for these parameters and statistics among the three methods.

\begin{table}[]
 \centering
 \caption{Summary of the SAUC estimates under the true selection mechanism of $(c_1, c_0)=(1/\sqrt{2},1/\sqrt{2})$ given $p\approx 0.7$. The estimates are summarized by mean (standard error) over 100 simulated meta-analyses; the values are multiplied by 100.}
 \label{tab:ressimulate}
 \resizebox{\linewidth}{!}{
 \begin{tabular}{cccccc}
\hline \multirow{2}{*}{ Experiment } & \multirow{2}{*}{ Method } & \multirow{2}{*}{ True } & $S=15$ & $S=25$ & $S=50$ \\
\cline{4-6} & & & AVE(SD) & AVE(SD) & AVE(SD) \\
\hline \multirow{5}{*}{1} & MLE with published studies & \multirow{5}{*}{83.2} & $88.1(4.1)$ & 87.3(3.5) & 87.7(2.3) \\
 & Method of \cite{zhou2023likelihood} & & 85.0(4.2) & $85.0(3.3)$ & 85.0(2.5) \\
 & Proposal with $(c_1, c_0)=(1/\sqrt{2},1/\sqrt{2})$ & & $84.8(5.8)$ & 84.0(4.5) & 83.5(3.4) \\
 & Proposal with $(c_1, c_0)=(1,0)$ & & 87.0(4.7) & 86.2(3.8) & 86.5(2.6) \\
 & Proposal with $(c_1, c_0)=(0,1)$ & & 88.5(4.1) & 88.0(3.4) & 88.6(2.1) \\
\hline \multirow{5}{*}{2} & MLE with published studies & \multirow{5}{*}{79.8} & $81.1(7.4)$ & $83.0(5.2)$ & 83.4(3.6) \\
 & Method of \cite{zhou2023likelihood} & & 75.3(8.8) & $77.5(7.4)$ & $77.5(5.2)$ \\
 & Proposal with $(c_1, c_0)=(1/\sqrt{2},1/\sqrt{2})$ & & 78.2(8.4) & $80.2(6.6)$ & 79.9(4.9) \\
 & Proposal with $(c_1, c_0)=(1,0)$ & & 81.5(7.1) & $83.1(5.3)$ & $83.2(3.5)$ \\
 & Proposal with $(c_1, c_0)=(0,1)$ & & 81.5(7.1) & $82.9(5.4)$ & 83.3(3.7) \\
\hline \multirow{5}{*}{3} & MLE with published studies & \multirow{5}{*}{86.9} & 88.4(3.4) & 89.2(2.6) & 89.7(1.5) \\
& Method of \cite{zhou2023likelihood} & & 84.3(4.2) & $84.6(4.0)$ & $85.7(2.4)$ \\
 & Proposal with $(c_1, c_0)=(1/\sqrt{2},1/\sqrt{2})$ & & 87.3(4.3) & $87.9(3.8)$ & $88.4(2.7)$ \\
 & Proposal with $(c_1, c_0)=(1,0)$ & & 89.0(3.7) & $90.0(3.0)$ & $90.7(1.8)$ \\
 & Proposal with $(c_1, c_0)=(0,1)$ & & 88.5(3.3) & $89.2(2.6)$ & $89.7(1.5)$ \\
\hline \multirow{5}{*}{4} & MLE with published studies & \multirow{5}{*}{83.2} & 84.7(3.1) & 84.3(2.3) & 84.5(1.4) \\
 & Method of \cite{zhou2023likelihood} & & 82.2(3.3) & $81.3(2.3)$ & 81.3(1.4) \\
 & Proposal with $(c_1, c_0)=(1/\sqrt{2},1/\sqrt{2})$ & & $83.8(3.6)$ & $83.3(2.5)$ & 83.5(1.9) \\
 & Proposal with $(c_1, c_0)=(1,0)$ & & 84.5(3.3) & $84.1(2.3)$ & 84.4(1.5) \\
 & Proposal with $(c_1, c_0)=(0,1)$ & & 85.0(3.1) & $84.5(2.3)$ & 84.7(1.4) \\
\hline \multirow{5}{*}{5} & MLE with published studies & \multirow{5}{*}{$79.8$} & 81.0(4.2) & 81.1(3.3) & $81.2(2.1)$ \\
 & Method of \cite{zhou2023likelihood} & & $77.8(4.7)$ & $77.3(4.0)$ & $77.3(2.6)$ \\
 & Proposal with $(c_1, c_0)=(1/\sqrt{2},1/\sqrt{2})$ & & $79.8(5.4)$ & 80.0(4.4) & $80.1(3.1)$ \\
 & Proposal with $(c_1, c_0)=(1,0)$ & & 81.2(4.3) & 81.2(3.5) & 81.1(2.2) \\
 & Proposal with $(c_1, c_0)=(0,1)$ & & 81.2(4.3) & $81.1(3.3)$ & 81.2(2.1) \\
\hline \multirow{5}{*}{6} & MLE with published studies & \multirow{5}{*}{86.9} & $87.2(2.3)$ & 87.4(2.2) & 87.6(1.2) \\
 & Method of \cite{zhou2023likelihood} & & 83.9(3.7) & 84.3(2.6) & 84.3(1.7) \\
 & Proposal with $(c_1, c_0)=(1/\sqrt{2},1/\sqrt{2})$ & & $86.8(3.3)$ & $87.5(2.9)$ & 87.8(1.9) \\
 & Proposal with $(c_1, c_0)=(1,0)$ & & $87.8(2.3)$ & $88.1(2.4)$ & 88.5(1.5) \\
 & Proposal with $(c_1, c_0)=(0,1)$ & & 87.2(2.3) & $87.4(2.2)$ & 87.6(1.2) \\
\hline
\end{tabular}
 }
 
\end{table}

\section{Discussions}
\label{ss:discussion}

We proposed a sensitivity analysis to address PB in meta-analysis of diagnostic studies. Most existing sensitivity analysis methods rely on the normal approximation for the pair of empirical logit-transformed sensitivities and specificities, we utilize the bivariate binomial model that uses exact within-study binomial model. The bivariate binomial model is more suitable for sparse data cases and possesses better finite sample performance. 
To our best knowledge, \cite{hattori2018sensitivity} is the only sensitivity analysis method applicable to the bivariate binomial model. \cite{hattori2018sensitivity} used the Heckman-type selection function. It described a selective publication process under which studies of larger sample size and of larger AUC were more likely to be published. On the other hand, the method for the SROC was designed to make inference based on the observations of pairs of sensitivity and specificity, and some studies might not report the AUC. Thus, publication of such studies would be determined by sensitivity and specificity rather than the AUC.  To address this issue, we extended the Copas t-statistics selection model to the meta-analysis of diagnostic studies with the bivariate binomial model. 
The selection function defines selective publication by the observed statistic of each study and its cut-off value. Such selective publication based on observed quantities would be more appealing. Since the true structure of the selective publication is unknown, we cannot determine which selection model would be more relevant between the Copas-Heckman and Copas t-statistics selection model. It is important to evaluate robustness against various kinds of potential selective publication processes. Thus, our development adds a useful tool to evaluate the potential impacts of selective publication processes alternative to the Copas-Heckman selection model. The results of both the real applications and simulation studies showed the applicability of our proposal.

We consider the selection function as a monotone function with the t-statistics of the linear combination of logit-transformed sensitivity and specificity. Our methods can reflect several publication mechanisms including those determined by lnDOR, sensitivity, and specificity. However, the true underlying publication mechanism is not easy to identify or even verify with limited information from only published studies. Thus, we recommend to conduct comprehensive sensitivity analysis with multiple selection functions corresponding to various t-statistics in practice.

We took a sensitivity analysis approach with the marginal selection probability $p$ fixed. In reality, $p$ is unknown. Thus, we need to consider several values of the marginal selection probability. \cite{li2021diagnostic} proposed EM algorithms to quantitatively estimate the parameters in the bivariate normal model and the SROC curve/SAUC, adjusting for PB without assuming a given selection probability. Their methods are based on the Copas-Heckman selection model. It would be interesting to develop methods to adjust for PB in meta-analysis of diagnostic studies with the Copas t-statistics selection functions, without specifying the sensitivity parameters and the marginal selection probability.

% \section*{Software}
% \label{software}

% Software in the form of R together with a sample input data set and complete documentation is available on GitHub at \url{https://github.com/Taojun-Hu/meta-analysis-pb-diagnostic}.

\section*{Acknowledgements}
% The authors would like to thank the editor, associate editor, and reviewers for their helpful and insightful comments.
This research was partly supported by Grant-in-Aid for Challenging Exploratory Research (16K12403) and for Scientific Research (16H06299, 18H03208) from the Ministry of Education, Science, Sports and Technology of Japan.

\section*{Supplementary Materials}
Web Appendices, referenced in Sections~\ref{ss:glmms}, Sections~\ref{ss:pb}, and \ref{ss:realanalysis}, are available with this paper at the Biometrics website on Wiley Online Library.
\vspace*{-8pt}

\bibliographystyle{biom} 
\bibliography{biomtemplate}

\label{lastpage}

\end{document}